\documentclass{elsart} 
\usepackage{epsfig, amssymb} 
\begin{document} 
 
\begin{frontmatter} 
 
\title{Level dynamics in pseudointegrable billiards: an experimental study} 
 
\author{Yuriy Hlushchuk$^1$}, 
\ead{yuriy.hlushchuk@physik.uni-giessen.de} 
\author{Ulrich Kuhl$^2$}, 
\ead{ulrich.kuhl@physik.uni-marburg.de} 
\author{Stefanie Russ$^1$} 
\ead{stefanie.russ@physik.uni-giessen.de} 
 
\address{$^1$Institut f\"ur Theoretische Physik III, Universit\"at Giessen, \\ D-35392 Giessen, Germany} 
\address{$^2$FB Physik, Universit\"at Marburg, \\ D-35032 Marburg, Germany}

\begin{abstract} 
The level dynamics of pseudointegrable systems with different genus numbers $g$
is studied experimentally using microwave cavities. 
For higher energies the distribution of the eigenvalue velocities is Gaussian,
as it is expected for chaotic systems with time-reversal symmetry, and shows no
dependence on $g$. 
Also the curvature distribution $P(k)$ for large $k$ is decaying as it is expected for chaotic systems, i.e. $P(k) \sim |k|^{-3}$. 
For small $k$ an intermediate behavior is found, where $P(k)$ changes from integrable towards chaotic behavior with growing $g$. 
\end{abstract} 
 
\end{frontmatter}

\section{Introduction} 
 
The motion of a classical particle in a billiard system can show regular, chaotic or intermediate behavior, depending on the billiard geometry. 
A potential well of the same geometry as the corresponding classical billiard -- a quantum billiard -- reflects this regular, chaotic or intermediate behavior in the statistics and the dynamics of its eigenvalues. 
From spectra only static correlations like nearest neighbor distributions can be obtained \cite{Mehta}. 
Level dynamics additionally observes eigenvalues in dependence of an external parameter. 
 
Pechukas \cite{Pec83} and Yukawa \cite{Yukawa85} established the formal analogy between a one-dimensional classical gas and a perturbed quantum system 
$H=H_0+tV$,
where $H_0$ is the Hamiltonian of the unperturbed system, and $tV$ is a perturbation. 
Here, the change of the eigenvalues $E_n(t)$ with the parameter $t$ is 
governed by the same ''equations of motion'' as the positions $x_n(t)$ of the gas particles propagating with time $t$. 
In this picture, a chaotic system can be described as a one-dimensional gas of particles with repulsive long-range interactions between them.
An integrable system, on the other hand, refers to an ideal gas with no interaction or one close to zero, when tunneling effects are taken into account \cite{braun2001}. 
The same equations of motion as for the Pechukas-Yukawa gas have been derived for the billiard system under the shift $\Delta l$ of a billiard wall \cite{Kollmann94,Stoeckmann97}, 
where $\Delta l$ (or the length of the billiard $l$) refers to the time $t$.

In this paper, we are interested in the level dynamics of two-dimensional pseudointegrable billiards \cite{richens}. 
Examples are polygons with only rational angles $n_i\pi/m_i$, with $n_i, m_i\in \mathbb{N}$ and at least one $n_i>1$. 
They are characterized by their genus number 
$g = 1 + (M/2) \sum_{i=1}^{J}(n_i-1)/m_i$, 
where $J$ is the number of angles and $M$ is the least common multiple of the $m_i$. 
The level statistics of pseudointegrable systems has been found intermediate between those of integrable and chaotic systems \cite{Cheon89,Shudo92,Shudo93,Shudo94,Russ01,Hlushchuk03,Wiersig03}. 
In the following, we concentrate on their level dynamics, i.e.\ on the distribution of the velocities and curvatures of the eigenvalues $E_n(l)$. 
 
It has been derived in \cite{Yukawa85} that the distribution of the velocities $v_n = d E_n/ d l$ of chaotic systems is Gaussian, 
$P(v)=\sqrt{\beta \over 2 \pi} \exp{(-{\beta \over 2}v^2)}$. 
The level curvature is defined as the second derivative of the eigenvalues
$K_n(l) = d^2{E}_n(l)/dl^2$.
We scale it to the dimensionless curvature
\begin{equation}\label{ScalingOfCurv} 
k={K \Delta \over \pi <(dE_n/dl)^2>},
\end{equation}
where $\Delta$ denotes the average level spacing \cite{Zakrzewski93,vonOppen94,vonOppen95}.
The trajectories of particles without interaction are straight lines, i.e., all $k_n (l) = 0$, which corresponds to $P(k)\sim\delta(0)$ for integrable systems. 
In a chaotic billiard, on the other hand, the interaction between the ''eigenvalue particles'' are long-ranged, leading to curved trajectories with a broad distribution $P(k)$. 
The curvatures $k$ for pure random Gaussian orthogonal ensembles (GOE) \cite{Mehta} obey the generalized Cauchy distribution \cite{Zakrzewski93,vonOppen94,vonOppen95} 
\begin{equation}\label{ScalCurvDistrEq} 
P(k)= {1 \over 2} {1 \over (1+k^2)^{3/2}}. 
\end{equation} 
Eq.(\ref{ScalCurvDistrEq}) corresponds to a power-law behavior for large $k$, $P(k) \sim |k|^{-3}$, in agreement with earlier derivations \cite{Gaspard}. 
The small curvature behavior has been found to be non generic, when families of periodic orbits like bouncing balls are not taken explicitely into account \cite{Zakrzewski93,Takami92} (see below). 
 
Numerical works on pseudointegrable billiards, as the barrier billiard \cite{wiersig} and systems approaching the chaotic Sinai billiard by polygons with different genus numbers between $g=2$ and $g=7$ \cite{Simmel95} 
found that the trajectories of the eigenvalues do not cross each other (as in integrable billiards), but show ''avoided crossings'', indicating that some repulsive interaction must be present. 
In \cite{Bogomolny}, a model of short-range interactions was used to 
calculate the spectral statistics of pseudointegrable systems. 
In \cite{Simmel95}, the curvature distribution $P(K)$ was calculated, which showed the expected GOE behavior for large $K$. Supprisingly, a peak was found for $|K|$ close to zero that was larger in the case of the Sinai billiard than for the polygon billiards.
As also pointed out by the authors, this demands further investigation and we will show that it might be due to the fact that the bouncing ball orbits between parallel walls were not taken into account explicitely.
 
\section{Systems and Experiment} 

\unitlength 1.0mm 
\vspace*{0mm} 
\begin{figure} 
\begin{center} 
\begin{picture}(130,50) 
\def\epsfsize#1#2{0.55#1} 
\put(0,5){\epsfbox{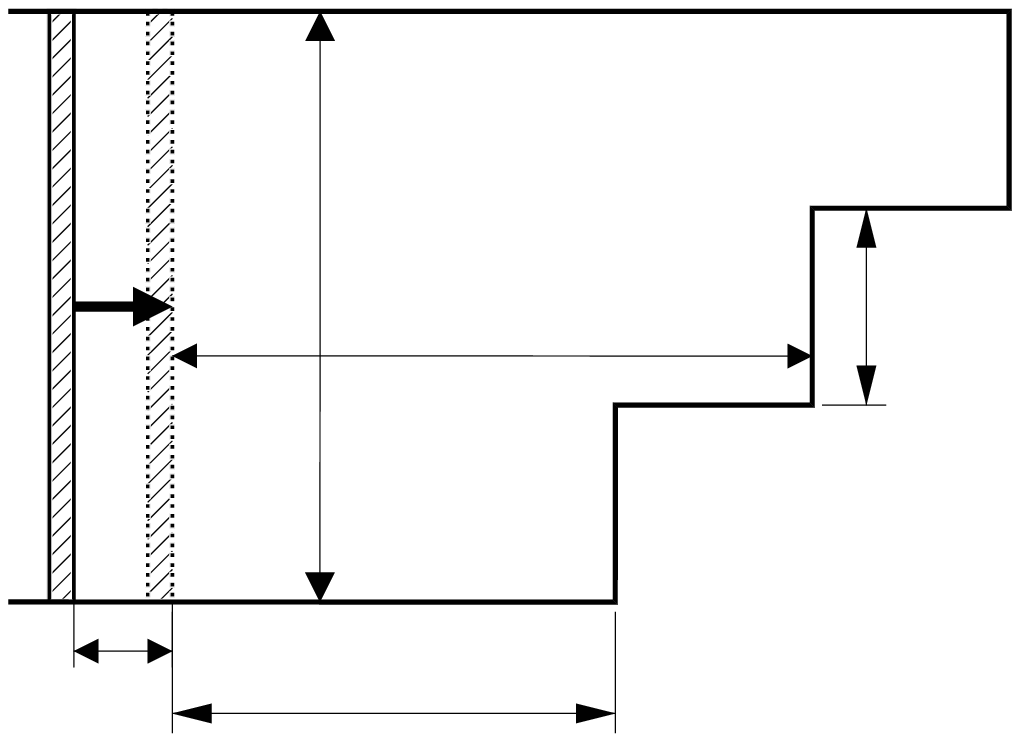}} 
\put(70,8){\epsfbox{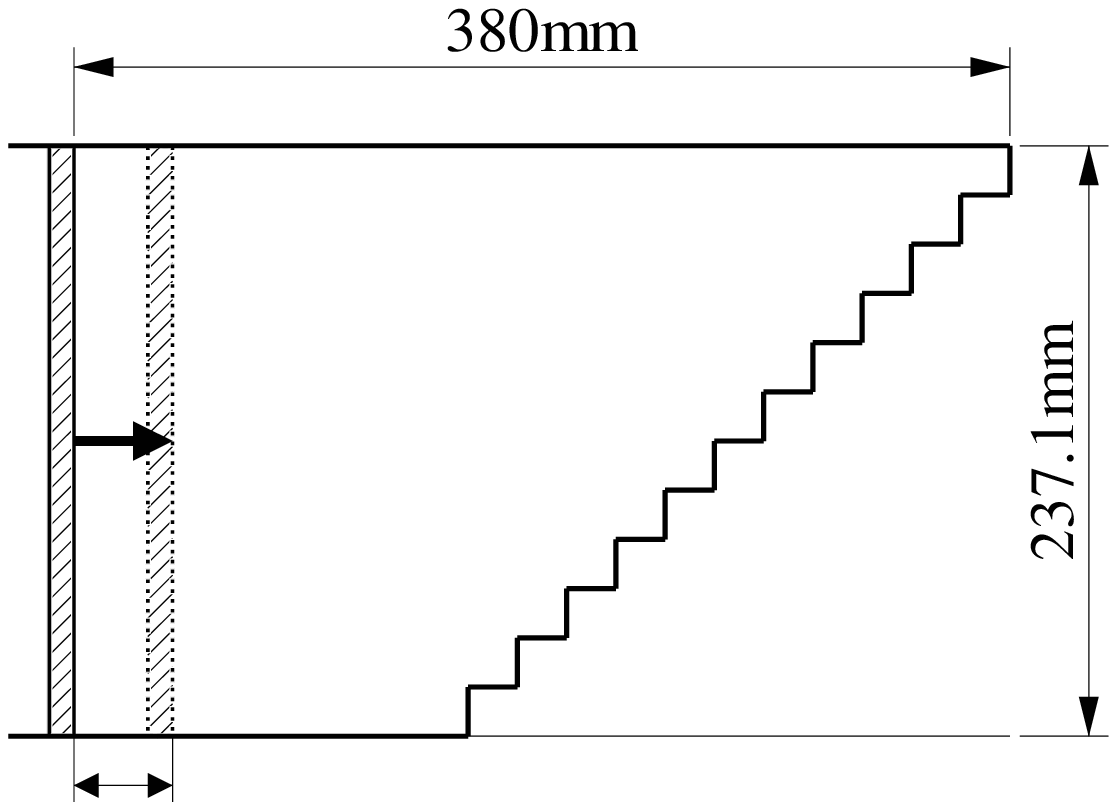}} 
\put(7.5,6.5){\makebox(1,1){{\small $\Delta l$ }}} 
\put(23,8.5){\makebox(1,1){{\small $w_{bbo,i}$ }}} 
\put(24,36){\makebox(1,1){{\small $\ell_{bbo,i}$ }}} 
\put(55,29){\makebox(1,1){{\small $w_{bbo,j}$ }}} 
\put(33,29){\makebox(1,1){{\small $\ell_{bbo,j}$ }}} 
\put(78,6.5){\makebox(1,1){{\small $\Delta l$ }}} 
\put(48,4){\makebox(1,1){{\bf (a) }}} 
\put(118,4){\makebox(1,1){{\bf (b) }}} 
\end{picture} 
\end{center} 
\caption[]{\small  Investigated billiards with genus $g=3$ and $g=12$. 
One wall was shifted in steps of $0.25 \, {\rm mm}$. 
Some of the widths $w_{bbo}$ and the length $\ell_{bbo}$ of the main bouncing ball orbits, taken into account for the unfolding of the spectra, are shown.} 
\label{FigBilliards} 
\end{figure} 
  
In this work, we measure experimentally the eigenvalues $E_n$ of pseudointegrable microwave billiards under the shift of a wall \cite{Stoeckmann90}. 
We investigate two pseudointegrable systems with genus $g=3$ and $12$ (see Fig.~\ref{FigBilliards}). 
The widths of both systems are equal $237.1 \, {\rm mm}$ and the initial lenghts are $l_0=380 \, {\rm mm}$. 
The length of the system $l$ can be changed by shifting one wall, which is used as the external parameter. 
The wall was moved in steps of $0.25 \, {\rm mm}$. 
For each position of the wall a spectrum from $0.5 \,{\rm GHz}$ up to $18.1 \, {\rm GHz}$ was measured. 
In total we have $161$ measurements giving $4{\rm cm}$ as a total change of length. 
A detailed analysis of all obtained spectra was performed up to a frequency of $15 \, {\rm GHz}$, yielding about $400$ resonances for each billiard. 
For higher frequencies the resonances could not be distinguished with sufficient reliability. 
 
In order to perform a statistical analysis of the given eigenvalue sequences independently from the change of size, the measured spectrum has to be unfolded to a mean level spacing of 1. 
To this end we used the Weyl formula \cite{Weyl} with the additional correction terms $N^{bbo}_i(E)$ that bring into account the existence of bouncing ball orbits (bbo) between parallel segments of the system, 
\begin{equation}\label{WeylEq} 
\varepsilon_n = N(E_n) = {A \over 4 \pi} E_n - {B \over 4 \pi} \sqrt{E_n} + \sum_{i} N^{bbo}_i(E_n),
\end{equation} 
where $A$ is the area and $B$ the perimeter of the billiard. 
The correction term derived in \cite{Sieber93} 
\begin{equation}\label{NbboEq} 
N^{bbo}_i(E_n) = {w_{bbo,i} \over {(2 \pi)}^{3/2}} \sqrt{2 \over \ell_{bbo,i}} E_n^{1/4} \sum_{m=1}^{\infty} m^{-3/2} \cos (2 m \ell_{bbo,i} \sqrt{E_n} - {3 \pi \over 4} ), 
\end{equation} 
is used additively for every parallel segment in our systems. 
Here $w_{bbo,i}$ is the width of the $i$-th parallel part of the billiard boundary and $\ell_{bbo,i}$ is the distance between its opposite walls (see also Fig.~\ref{FigBilliards}). 
From now on, we refer to the unfolded energy values $\varepsilon_n$. 
 
\section{Results} 

\unitlength 1.0mm 
\vspace*{0mm} 
\begin{figure} 
\begin{picture}(0,45) 
\def\epsfsize#1#2{0.6#1} 
\put(5,5){\epsfbox{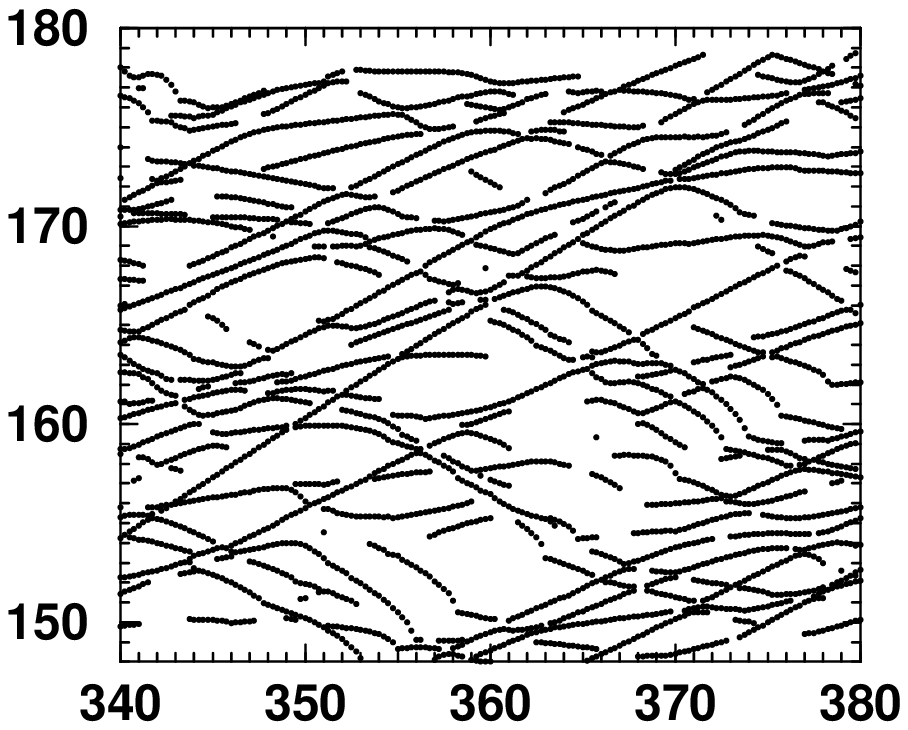}} 
\put(75,5){\epsfbox{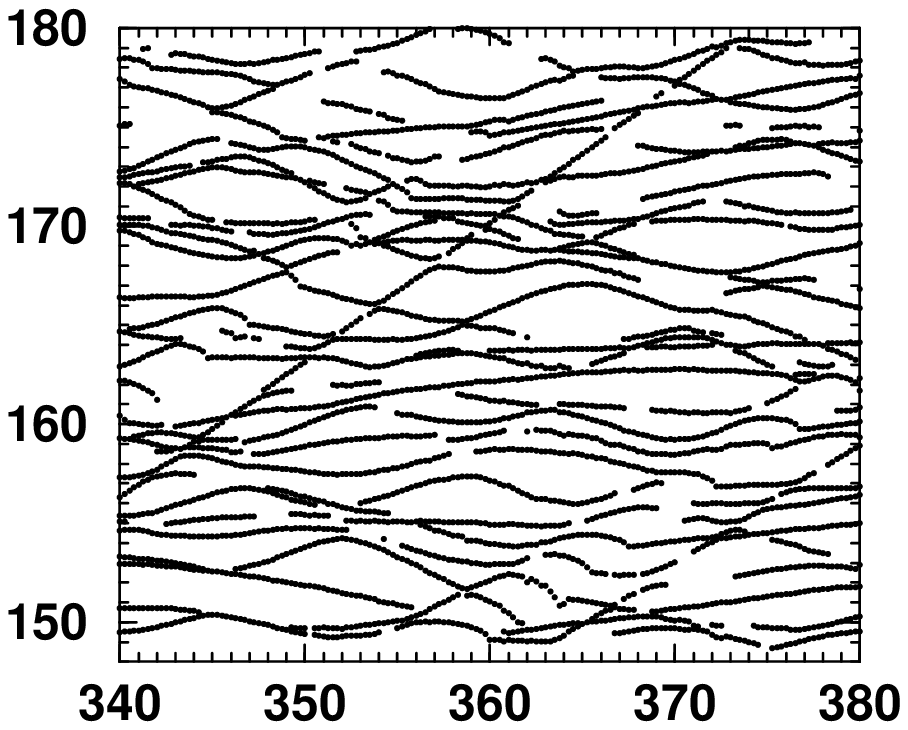}} 
\put(3,29){\makebox(1,1){{ \large $\varepsilon_n$ }}} 
\put(73,29){\makebox(1,1){{ \large $\varepsilon_n$ }}} 
\put(45,0){\makebox(1,1){{ \small $l$ [mm] }}} 
\put(115,0){\makebox(1,1){{ \small $l$ [mm] }}} 
\put(63,42){\makebox(1,1){{ \bf (a) }}} 
\put(133,40){\makebox(1,1){{ \bf (b) }}} 
\end{picture} 
\caption[]{\small The unfolded energy levels $\varepsilon_i$, $i =
151,\dots,180$ as a function of the billiard length $l$ for the $g=3$ -system (a) and the $g=12$ -system (b). 
It can be recognized that there are more straight lines in the spectrum of the billiard with $g=3$.} 
\label{fig_spagetti} 
\end{figure}  

In Fig.~\ref{fig_spagetti} we present a part of the unfolded energy spectra $\varepsilon_n$ as a function of the length $l$ of the billiard for both systems. 
Both figures look quite similar, though the amount of straight lines seems to be higher for the $g=3$ -system than for the $g=12$ -system, which could be due to a shorter range of the interactions between particles in pseudointegrable systems. 
In the following, we investigate this assumption quantitatively by calculating the velocity and the curvature distribution functions. 
 
In the case of GOE systems \cite{Pec83,Yukawa85}, one expects a Gaussian distribution for the velocity distribution $P(\tilde{v})$ of the scaled velocity $\tilde{v} =
(d \varepsilon/d l) / <{(d \varepsilon/ d l)}^2>^{1/2}$,  
$P(\tilde{v}) = (1 / \sqrt{2 \pi}) \exp{(-\tilde{v}^2 / 2)}$. 
However, for low-lying energies, the statistical properties often deviate from theoretical expectations.
Therefore we investigated $P(\tilde{v})$ for the low-energy and the high-energy regime separately. Fig.~\ref{fig_velocities}~(a,~b) shows the high-energy regime, $\varepsilon \in [350, 400]$, which is for both systems close to the Gaussian curves of chaotic systems (whereas large deviations occur at low energies). 

\unitlength 1.0mm 
\vspace*{0mm} 
\begin{figure} 
\begin{picture}(0,40) 
\def\epsfsize#1#2{0.30#1} 
\put(15,2){\epsfbox{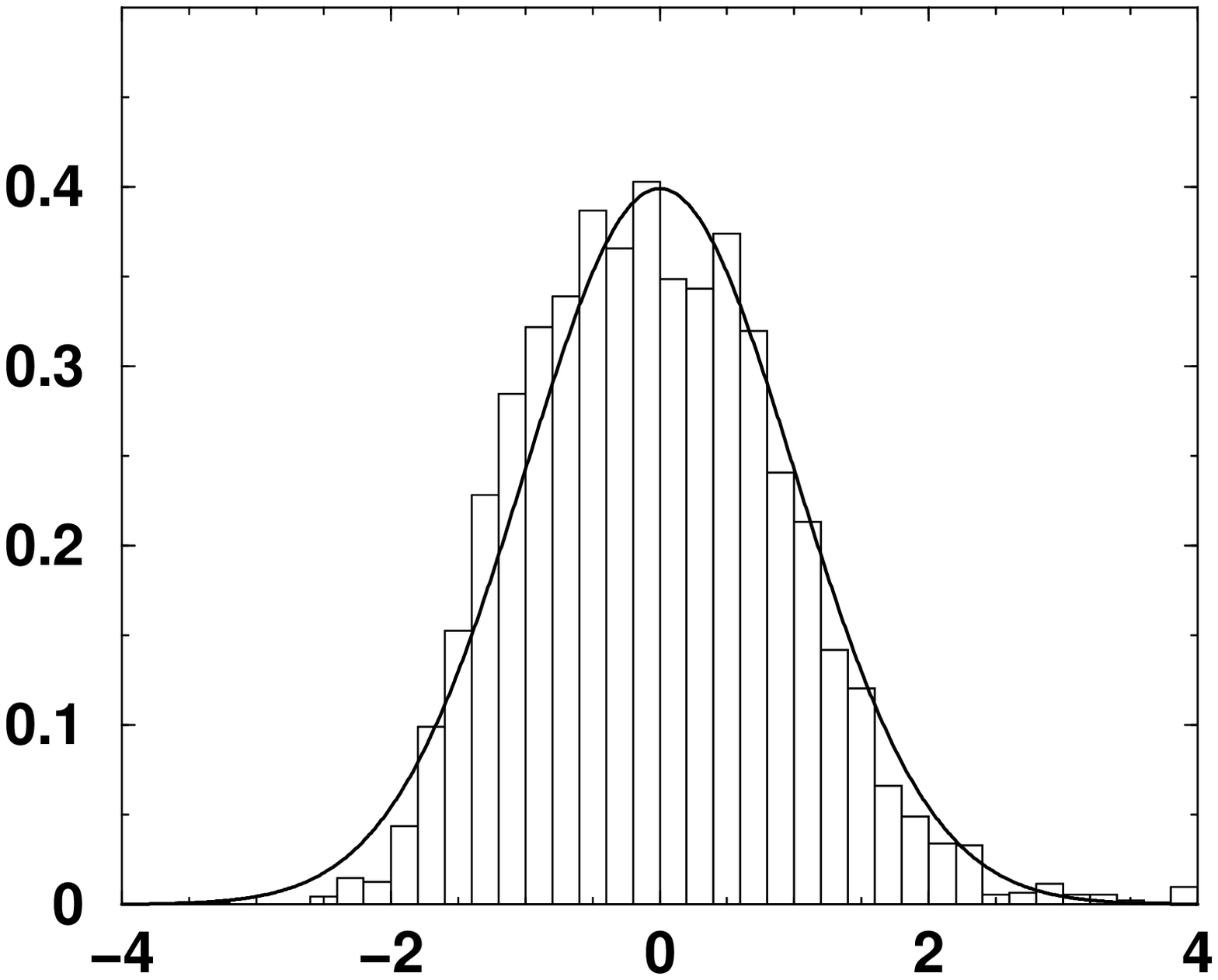}} 
\put(80,2){\epsfbox{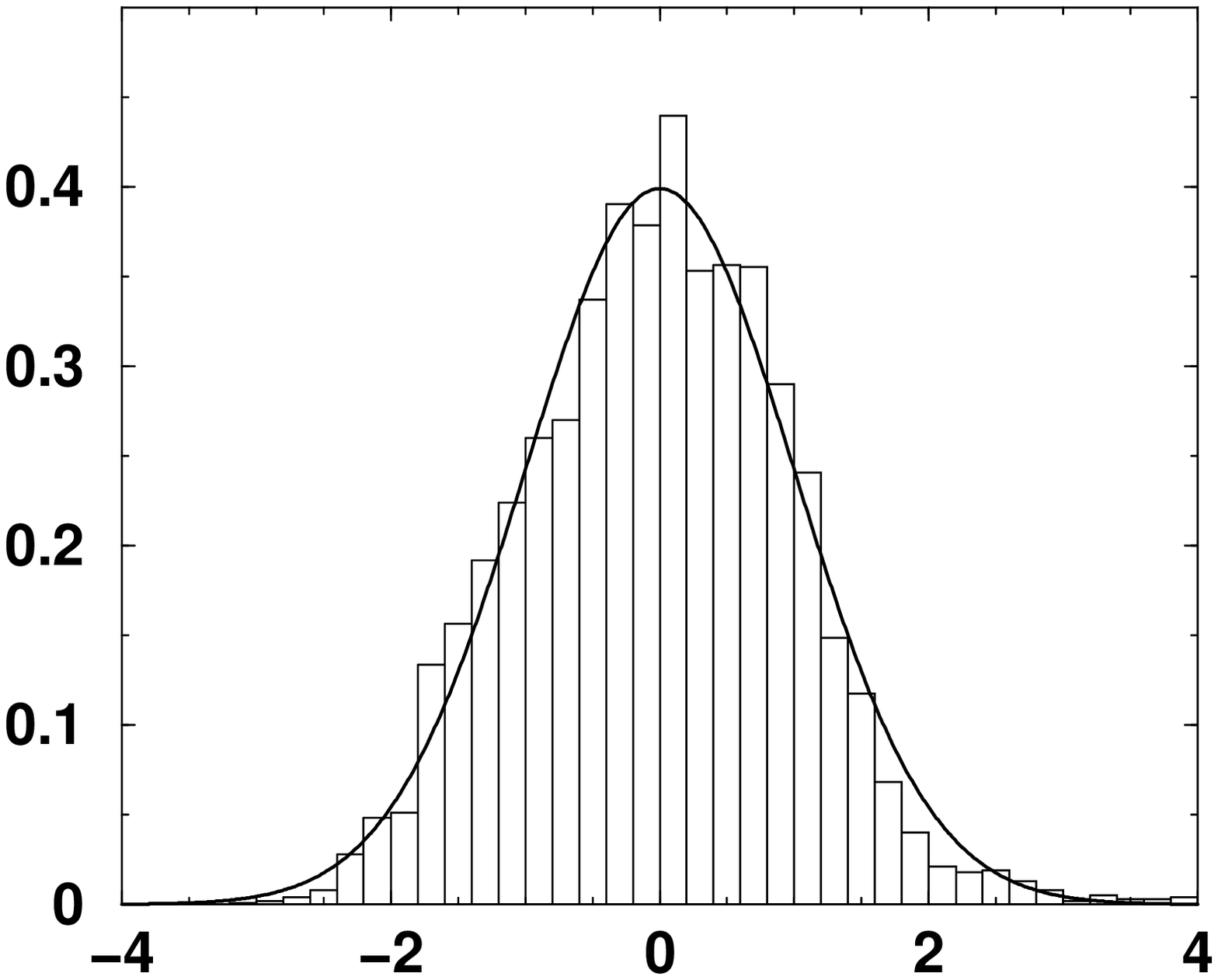}} 
\put(12,39){\makebox(1,1){{ $P(\tilde{v})$ }}} 
\put(77,39){\makebox(1,1){{ $P(\tilde{v})$ }}} 
\put(57,0){\makebox(1,1){{ $\tilde{v}$ }}} 
\put(122,0){\makebox(1,1){{ $\tilde{v}$ }}} 
\put(55,33){\makebox(1,1){{ \bf (a) }}} 
\put(120,33){\makebox(1,1){{ \bf (b) }}} 
\end{picture} 
\caption[]{\small  Distribution $P(\tilde{v})$ of the rescaled velocities $\tilde{v}$ for systems with $g=3$ and $g=12$\, in the high energy region ($\varepsilon \in [350, 400]$). 
The solid line is a Gaussian.} 
\label{fig_velocities} 
\end{figure}

For studying the curvatures $k$ we chose the same energy interval $\varepsilon \in [350, 400]$ as in Fig.~\ref{fig_velocities}. Fig.~\ref{fig_kruemm} (a,~b) shows $P(k)$ for large $k$ in a double-logarithmic scale for $g=3$ and $g=12$. 
Both experimental results show a $|k|^{-3}$ behavior of the tail (see solid line), which is expected for chaotic systems (GOE). 
Therefore, it seems that the decay is not depending on the genus number. 
Similar results were already reported for the pseudointegrable billiards of \cite{Simmel95}. 
This can be understood as follows: 
The large curvature behavior is mainly governed by the curvatures at the 
close encounters i.e.\,the avoided crossings. 
There, the curvature only depends on the interaction of the eigenvalue with its nearest neighbor. 
At short distances, the exact range of the interaction is unimportant and already short-range interactions are sufficient to lead to the same $|k|^{-3}$ behavior as in GOE systems. 

\unitlength 1.0mm 
\vspace*{0mm} 
\begin{figure} 
\begin{picture}(0,50) 
\def\epsfsize#1#2{0.4#1} 
\put(8,4){\epsfbox{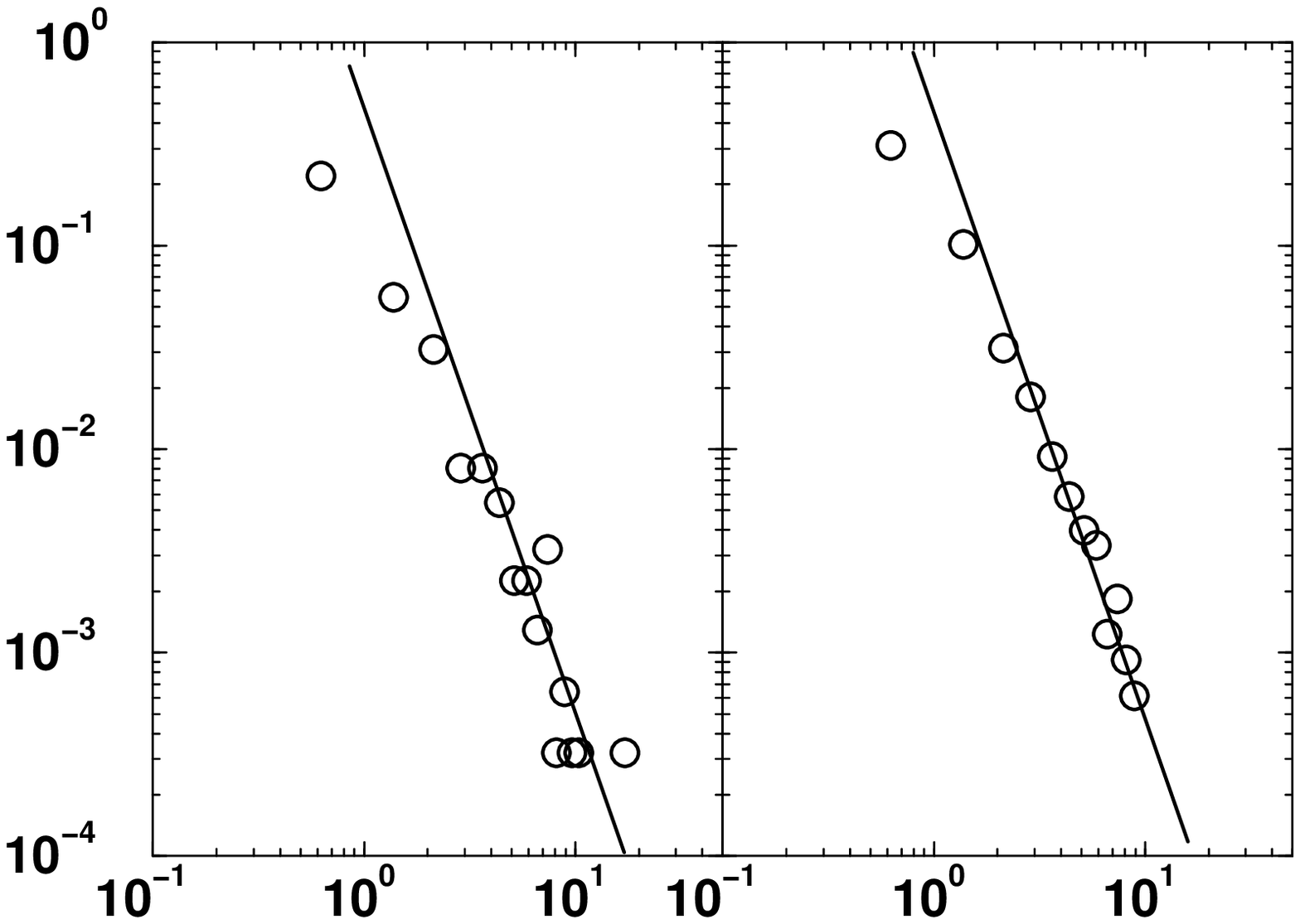}} 
\def\epsfsize#1#2{0.295#1} 
\put(84,4.5){\epsfbox{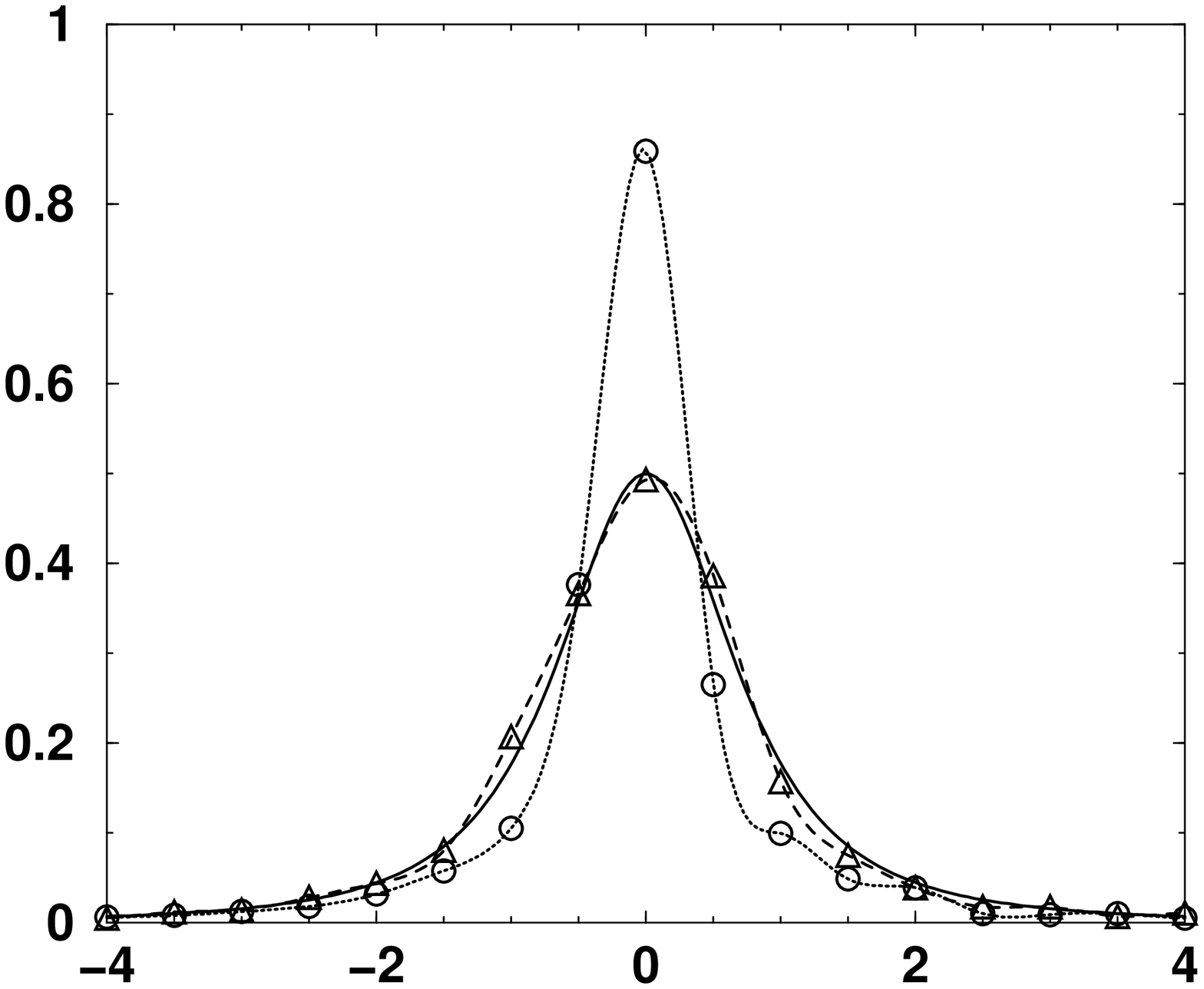}} 
\put(5,44){\makebox(1,1){{\small $P(|k|)$ }}} 
\put(81,44){\makebox(1,1){{\small $P(k)$ }}} 
\put(40,1){\makebox(1,1){{\small $|k|$ }}} 
\put(68,1){\makebox(1,1){{\small $|k|$ }}} 
\put(119,1){\makebox(1,1){{\small $k$ }}} 
\put(36,42){\makebox(1,1){{ \bf (a) }}} 
\put(65,42){\makebox(1,1){{ \bf (b) }}} 
\put(129,42){\makebox(1,1){{ \bf (c) }}} 
\end{picture} 
\caption[]{\small Distribution of the absolute value of the scaled curvatures $|k|$ in a double logarithmic scale for the $g=3$ -billiard (a) and the $g=12$ -billiard (b). 
The solid line indicates a $|k|^{-3}$ behavior as expected for GOE systems. 
Case (c) shows the distribution of $k$ for $g=3$ (circles) and $g=12$ (triangles) for $\varepsilon \in [350,400]$. 
The solid line shows the theoretical prediction for a GOE system (see Eq.~(\ref{ScalCurvDistrEq})). 
} 
\label{fig_kruemm} 
\end{figure}  

The distributions of small values of $k$, on the other hand, may depend on the range of the interactions.
In Fig.~\ref{fig_kruemm}~(c), we therefore show $P(k)$ on a linear scale and compare it to the theoretical prediction for a GOE system (solid line, cf. Eq.~(\ref{ScalCurvDistrEq})).
Whereas for the $g=3$-system, the values of $P(k)$ at small $k$ are much higher, $P(k)$ of the $g=12$-billiard is indistinguishable from Eq.~(\ref{ScalCurvDistrEq}).
This corresponds to the observation of more straight parts of eigenvalue trajectories in the $g=3$-system (see Fig.~\ref{fig_spagetti}) and indicates that $P(k)$ changes from a $P(k)\sim\delta(0)$ for integrable systems towards Eq.~(\ref{ScalCurvDistrEq}) with growing $g$. 
 
\section{Summary and Conclusions} 
 
In summary, we investigated experimentally the level dynamics of two pseudointegrable billiards with different genus numbers $g$ and calculated the distribution $P(\tilde{v})$ of the velocities and $P(k)$ of the curvatures of the eigenvalues. 
While the velocity distribution as well as $P(k)$ for large $k$ shows the
behavior expected for GOE systems, $P(k)$ depends on $g$.
Using an unfolding procedure that takes the contributions of the bouncing ball orbits between parallel segments of the system into account, we found that $P(k)$ approaches Eq.~(\ref{ScalCurvDistrEq}) of the GOE system with increasing $g$.
It will be very interesting to investigate also systems with $3 < g < 12$ as well as with other system details, to see if this behavior is general.
 
The distribution $P(k)$ for small curvatures does not correspond to the results presented in \cite{Simmel95}, which could be due to the different unfolding procedure. 
Contrary to our work, the unfolding by \cite{Simmel95} did not take into account the bouncing ball orbits between parallel walls and it is not evident what governs the behavior of the small curvatures in this case. 

Our results (see Fig.~\ref{fig_kruemm}~(c)) support the model of short-range interactions between eigenvalue particles in pseudointegrable systems. 
Small curvatures arise from eigenvalues in large distances from each other. 
When the interaction between them decays rapidly, eigenvalues beyond a certain distance do not influence each other and their trajectories become nearly straight lines. 
Our results indicate that the range of the interaction increases with $g$. 
 
We thank H.-J.~St\"ockmann and R.~Sch\"afer for valuable discussions and 
the Deutsche Forschungsgemeinschaft for financial support.

\end{document}